\documentclass{elsart5p}

\usepackage{natbib}

\usepackage{graphicx}
\usepackage{color}

\usepackage{amssymb}
\usepackage{amsmath}
\usepackage{listings}
\newcommand{\bm}[1]{\boldsymbol  #1 }
\newcommand{\dm}{{\rm d}}

\begin{document}

\begin{frontmatter}



\title{$N$-body simulation for self-gravitating collisional systems
  with a new SIMD instruction set extension to the x86 architecture,
  Advanced Vector eXtensions}


\author[UTsukuba]{Ataru Tanikawa\corauthref{cor1}},
\ead{tanikawa@ccs.tsukuba.ac.jp}
\author[UTsukuba]{Kohji Yoshikawa},
\author[UTsukuba]{Takashi Okamoto} \&
\author[UTsukuba]{Keigo Nitadori}

\address[UTsukuba]{Center for Computational Science, University of Tsukuba, 1--1--1, Tennodai, Tsukuba, Ibaraki 305--8577, Japan}

\corauth[cor1]{Corresponding author.}

\begin{abstract}
  We present a high-performance $N$-body code for self-gravitating
  collisional systems accelerated with the aid of a new SIMD
  instruction set extension of the x86 architecture: Advanced Vector
  eXtensions (AVX), an enhanced version of the Streaming SIMD
  Extensions (SSE). With one processor core of Intel Core i7-2600
  processor (8MB cache and 3.40 GHz) based on Sandy Bridge
  micro-architecture, we implemented a fourth-order Hermite scheme
  with individual timestep scheme \citep{Makino92}, and achieved the
  performance of $\sim 20$ giga floating point number operations per
  second (GFLOPS) for double-precision accuracy, which is two times
  and five times higher than that of the previously developed code
  implemented with the SSE instructions \citep{Nitadori06b}, and that
  of a code implemented without any explicit use of SIMD instructions
  with the same processor core, respectively. We have parallelized the
  code by using so-called NINJA scheme \citep{Nitadori06a}, and
  achieved $\sim 90$ GFLOPS for a system containing more than $N=8192$
  particles with 8 MPI processes on four cores. We expect to achieve
  about $10$ tera FLOPS (TFLOPS) for a self-gravitating collisional
  system with $N \sim 10^5$ on massively parallel systems with at most
  $800$ cores with Sandy Bridge micro-architecture. This performance
  will be comparable to that of Graphic Processing Unit (GPU) cluster
  systems, such as the one with about 200 Tesla C1070 GPUs
  \citep{Spurzem10}. This paper offers an alternative to collisional
  $N$-body simulations with GRAPEs and GPUs.
\end{abstract}

\begin{keyword}
Stellar dynamics \sep Method: $N$-body simulations

\end{keyword}

\end{frontmatter}

\section{Introduction}

$N$-body simulations are a powerful tool to follow dynamical evolution
of self-gravitating many-body systems, such as planetary systems, star
clusters, galaxies, galaxy clusters, and the large scale structure in
the universe. In an $N$-body simulation, we solve the following
equation of motion of particles:
\begin{equation}
  \frac{\dm^2\bm{r}_i}{\dm t^2} = \sum_{j \neq i}^N \frac{Gm_j}{\left(|\bm{
      r}_{ij}|^2 + \varepsilon^2 \right)^{3/2}}\bm{r}_{ij},
       \label{eq:eom}
\end{equation}
where $G$ is the gravitational constant, $\varepsilon$ is the
softening length, $N$ is the number of particles, $m_i$ and $\bm{r}_i$
are respectively the mass and position of $i$th particle, and
$\bm{r}_{ij}=\bm{r}_j - \bm{ r}_i$. For self-gravitating collisional
systems such as planetary systems and star clusters, one needs to keep
sufficiently good accuracy and hence resort to a brute-force scheme, a
direct scheme to compute gravitational forces. In fact, the number of
particles adopted in recent numerical simulations of these systems is
not always large enough. For example, the number of stars in a single
globular cluster is typically $\sim 10^6$, whereas the number of
particles in recent numerical simulation of globular clusters is $\sim
10^5$ at most. This is the reason why efficient ways to compute
gravitational force for a given set of particles are intensively
studied in many aspects.

The use of external hardwares is one of the most popular ways to
compute gravitational force very quickly. GRAPEs
\citep{Ito91,Makino97,Makino03}, which are special purposed
accelerators for $N$-body simulations, have greatly contributed to
$N$-body simulations of collisional systems. Recently,
\citet{Nitadori09} and \citet{Gaburov09} explored the capability of
commodity Graphics Processing Units (GPUs) as hardware accelerators
for $N$-body simulation and achieved a performance close to that of
GRAPE-DR \citep{Makino08}, the latest version of the GRAPE family.

Another approach to accelerate the computation of gravitational forces
is to utilize SIMD (Single Instruction, Multiple Data) instructions,
which can perform the same operation on multiple data simultaneously,
implemented on recent central processing units (CPUs) instead of
making use of external hardware accelerators. Since Pentium III
released by Intel Corporation in 1999, CPUs with x86 architecture have
supported a set of SIMD instructions called Streaming SIMD Extensions
(SSE). \citet{Nitadori06b} (NMH06) applied the SSE instructions to the
force calculation in $N$-body simulations, and calculated four
interactions among particles in parallel. NMH06 achieved about three
times higher performance than the case without the use of the SSE
instructions. Furthermore, \citet{Nitadori06a} (NMA06) demonstrated
that the performance of CPUs with the aid of the SSE instructions is
comparable to GRAPEs and GPUs in massively parallel simulations.

Recently, a new processor family with Sandy Bridge micro-architecture
has been released by Intel Corporation. The processor supports a new
set of instructions known as Advanced Vector eXtensions (AVX), an
enhanced version of the SSE instructions. In the AVX instruction set,
the width of the SIMD registers are extended from 128-bit to 256-bit.
Hence, an AVX instruction set is able to process twice amount of data
than the SSE instruction set. This suggests that a processor with the
AVX instruction set should be able to compute accelerations twice as
fast compared to its SSE-only counterpart. The AVX instructions are
also supported by the next-generation CPU "Bulldozer" released by AMD
Corporation.

In this paper, we present a new $N$-body code implemented using the
AVX instructions. In our $N$-body code, we adopt a fourth-order
Hermite scheme with individual timestep scheme \citep{Makino92} for
its time integration, which is widely used for collisional $N$-body
simulations. We have achieved $20$ giga floating point number
operations per second (GFLOPS) per processor core for $N$-body
simulations. This performance is two times higher than the case in
which we use the SSE instructions.

This paper is organized as follows. In section \ref{sec:algorithm}, we
outline the algorithm of the fourth-order Hermite scheme. In section
\ref{sec:implementation}, we describe our implementation for the
$N$-body code using the AVX instruction set. In section
\ref{sec:accuracy} and \ref{sec:performance}, we show an accuracy and
performance of our $N$-body code, respectively. The results are
summarized and discussed in section \ref{sec:summary}.

\section{Algorithm}
\label{sec:algorithm}

The fourth-order Hermite scheme is a kind of a predictor-corrector
method, developed by \cite{Makino92}. We outline its algorithm
below. The predictor of $i$th particle is given by Taylor series as
\begin{eqnarray}
  \bm{r}_i^{\rm (p)} &=& \bm{r}_i^{\rm (0)} + \bm{v}_i^{\rm (0)}
  {\Delta t}_i + \frac{\bm{a}_i^{\rm (0)}}{2} {{\Delta t}_i}^2 +
  \frac{\bm{j}_i^{\rm (0)}}{6} {{\Delta t}_i}^3 \label{eq:xpred}
  \\ \bm{v}_i^{\rm (p)} &=& \bm{v}_i^{\rm (0)} + \bm{a}_i^{\rm (0)}
     {\Delta t}_i + \frac{\bm{j}_i^{\rm (0)}}{2} {{\Delta
         t}_i}^2, \label{eq:vpred}
\end{eqnarray}
where ${\Delta t}_i$, $\bm{r}_i$, $\bm{v}_i$, $\bm{a}_i$, and $\bm{
  j}_i$ are the timestep, position, velocity, acceleration, and its
first-order time derivative (jerk) of $i$th particle, respectively,
and the superscripts $(0)$ and ${\rm (p)}$ indicate quantities at the
current time and predicted quantities, respectively. Using the
predicted position and velocity, $\bm{r}_i^{\rm (p)}$ and $\bm{
  v}_i^{\rm (p)}$, the acceleration and jerk of $i$th particle are
evaluated as
\begin{eqnarray}
  \bm{a}_i^{\rm (1)} &=& \sum_{i \neq j}^N \frac{G
    m_j}{\left[\left(r_{ij}^{\rm (p)}\right)^2 +
      \varepsilon^2\right]^{3/2}}\bm{r}_{ij}^{\rm (p)},
  \label{eq:acc} \\
  \bm{j}_i^{\rm (1)} &=& \sum_{i \neq j}^N
  \frac{Gm_j}{\left[\left(r_{ij}^{\rm (p)}\right)^2 +
      \varepsilon^2\right]^{3/2}} \left[\bm{v}_{ij}^{\rm (p)} -
    \frac{3\bm{r}_{ij}^{\rm (p)}\cdot\bm{v}_{ij}^{\rm
        (p)}}{\left(r_{ij}^{\rm (p)}\right)^2 + \varepsilon^2} \bm{
      r}_{ij}^{\rm (p)} \right], \label{eq:jrk}
\end{eqnarray}
where $\bm{r}_{ij}^{\rm (p)}=\bm{r}_j^{\rm (p)}-\bm{r}_i^{\rm (p)}$,
$\bm{v}_{ij}^{\rm (p)}=\bm{v}_j^{\rm (p)}-\bm{v}_i^{\rm (p)}$, and
$r_{ij}^{\rm (p)} = |\bm{r}_{ij}^{\rm (p)}|$, and the superscript
$(1)$ indicates quantities at the next time. Additionally, the
potential of $i$th particle, $\phi_i^{(1)}$, given by
\begin{equation}
  \phi_i^{\rm (1)} = - \sum_{i \neq j}^N
  \frac{Gm_j}{\left[\left(r_{ij}^{\rm
        (p)}\right)^2+\varepsilon^2\right]^{1/2}}, \label{eq:pot}
\end{equation}
is computed for checking the validity of the energy conservation.

The corrector is based on the third-order Hermite interpolation
constructed using the old acceleration and jerk, and the new ones
($\bm{a}_i^{\rm (0)}$, $\bm{j}_i^{\rm (0)}$, $\bm{a}_i^{\rm (1)}$, and
$\bm{j}_i^{\rm (1)}$, respectively). From the third-order Hermite
interpolation polynomial, we can obtain the second-order and
third-order time derivatives of acceleration, which are so-called snap
($\bm{s}_i^{\rm (0)}$) and crackle ($\bm{c}_i^{\rm (0)}$),
respectively. The formulas for the snap and crackle are expressed as
\begin{eqnarray}
  \bm{s}_i^{\rm (0)} &=& 2\left[-3\left(\bm{a}_i^{\rm (0)}-\bm{
      a}_i^{\rm (1)}\right)-\left(2\bm{j}_i^{\rm (0)}+\bm{j}_i^{\rm
      (1)}\right){\Delta t}_i\right] {\Delta t}_i^{-2}, \\ \bm{
    c}_i^{\rm (0)} &=& 6\left[2\left(\bm{a}_i^{\rm (0)}-\bm{
      a}_i^{\rm (1)}\right)+\left(\bm{j}_i^{\rm (0)}+\bm{j}_i^{\rm
      (1)}\right){\Delta t}_i\right]{\Delta t}_i^{-3}.
\end{eqnarray}
The corrections for the position and velocity are given by
\begin{eqnarray}
  \bm{r}_i^{\rm (1)} &=& \bm{r}_i^{\rm (p)} + \frac{\bm{s}_i^{\rm
      (0)}}{24} {{\Delta t}_i}^4 + \frac{\bm{c}_i^{\rm (0)}}{120}
  {{\Delta t}_i}^5 \\ \bm{v}_i^{\rm (1)} &=& \bm{v}_i^{\rm (p)} +
  \frac{\bm{s}_i^{\rm (0)}}{6} {{\Delta t}_i}^3 + \frac{\bm{
      c}_i^{\rm (0)}}{24} {{\Delta t}_i}^4.
\end{eqnarray}

Usually, individual timestep scheme is adopted in the Hermite
scheme. The timestep of $i$th particle is given by
\begin{equation}
  \Delta t_i = \eta \left(\frac{|\bm{a}_i^{\rm (0)}||\bm{s}_i^{\rm
      (0)}| + |\bm{j}_i^{\rm (0)}|^2}{|\bm{j}_i^{\rm (0)}||\bm{
      c}_i^{\rm (0)}| + |\bm{s}_i^{\rm
      (0)}|^2}\right)^{1/2}, \label{eq:timestep}
\end{equation}
where $\eta$ is an accuracy parameter \citep{Aarseth85}. Practically,
we employ a hierarchical timestep scheme \citep{Makino91} for SIMD
programing and parallelization; we discretize the timesteps in a power
of two hierarchy, where all timesteps are shorter or equal to the
timesteps obtained by equation (\ref{eq:timestep}), and are a power of
two subdivision of a predefined global timestep.

A detailed procedure of the Hermite scheme is as follows.
\begin{enumerate}
\item[1.] Select particles to be integrated (hereafter,
  $i$-particles), whose next times are nearest among all the
  particles.
\item[2.] Predict the positions and velocities of $i$-particles.
\item[3.] Predict the positions and velocities of particles which
  exert forces on $i$-particles (hereafter, $j$-particles).
\item[4.] Calculate accelerations, jerks, and potentials of
  $i$-particles.
\item[5.] Construct the Hermite interpolation for $i$-particles.
\item[6.] Correct the positions and velocities of $i$-particles.
\item[7.] Update $\bm{a}_i^{(0)}$ and $\bm{j}_i^{(0)}$ by substituting
  $\bm{a}_i^{(1)}$ and $\bm{j}_i^{(1)}$, respectively.
\item[8.] Calculate the next timesteps of $i$-particles.
\item[9.] Return to the step 1.
\end{enumerate}
The step 2 is a duplication of the step 3. This duplication has small
overhead, and is convenient when we perform this scheme in parallel.

\section{Implementation}
\label{sec:implementation}

We use the AVX instructions to implement force calculation in the
Hermite scheme, i.e. the step 4.  A cost for the force calculation is
order of $N^2$ and that for the other parts is order of $N$; thus the
force calculation dominates the total calculation time.  Below, we
describe the implementation of the force calculation.

\subsection{The SSE and AVX instructions}

Before we describe our implementation, let us briefly summarize the
SSE and AVX instructions and the difference between them.

The SSE instruction set is a SIMD instruction set introduced for the
first time in Pentium III processors to improve the performance of
media streaming, image processing and three-dimensional graphics by
executing $4$ single-precision (SP) or $2$ double-precision (DP)
floating-point operations in parallel.  Operations supported by the
SSE instruction set include addition, subtraction, multiplication,
division, square root, inverse square root, etc. In such operations,
dedicated registers with 128-bit length called ``XMM registers'' are
used to store the $4$ SP and $2$ DP floating-point numbers. Although
operations between data in the XMM registers and in main memory are
supported, those with all data stored in the XMM registers are
faster. The available number of the XMM registers in a single
processor core is 8 for x86 processors and 16 for x86\_64
processors. NMH06 utilized the SSE instruction set to accelerate the
force calculations in their collisional $N$-body code.

The AVX instruction set is an enhanced version of the SSE instruction
set, and there are two major differences between them. One is the
number of data on which an instruction is operated in parallel. The
length of the dedicated registers for the AVX instructions, ``YMM
registers'', is $256$ bits, two times longer than that of the XMM
registers. Thus, $8$ SP or $4$ DP floating-point operations can be
carried out simultaneously by the AVX instructions, while the SSE
instructions can execute $4$ SP floating-point operations or $2$ DP
floating-point operations in parallel. The lower $128$ bits of the YMM
registers are regarded as the XMM registers used by the SSE
instructions for backward compatibility. The number of the YMM
registers in one processor core is 16, equal to that of the XMM
registers.

The other is that the number of operands of most instructions has been
increased from two in the SSE instruction to three operands in the AVX
instructions. Two of the three operands are source operands, and the
other one is a destination operand, where the result of the operation
is stored.  Owing to this, while one of source operands is used as a
destination operand in the SSE instructions, and overwritten, we can
preserve both source operands at each AVX instruction, and use a
limited number of the YMM registers very efficiently.

Methods to implement the code explicitly using the SSE and AVX
instructions should be also addressed. In principle, if the compilers
were clever enough to detect any concurrent loop, it could generate
the code that effectively utilize these instructions. In reality,
however, the present-day compilers cannot fully resolve the dependency
among variables in the loops very well. Therefore, we have to
implement the SSE and AVX instructions explicitly using
assembly-languages or compiler-dependent intrinsic
functions. Intrinsic functions are supported by several compilers and
easy to program with. On the other hand, inline assembly-languages, we
can manually control the assignment of the XMM and YMM registers to
computational data, and minimize the access to the memory or the cache
memory by optimizing the use of individual registers. As already noted
above, calculations using SSE and AVX instructions are very efficient
if all the data are stored in the dedicated registers. Therefore, the
optimization of the use of the XMM and YMM registers is of crucial
importance. In this work, we adopt an implementation of the SSE and
AVX instructions using inline-assembly embedded in C-language. In the
followings, we only present implementations and results for x86\_64
processors with 16 XMM and YMM registers.

\subsection{Arithmetic precision}

In our implementation, we perform the force calculation in the
``mixed'' precision, in which, as is done in NMH06, only the relative
position vector between $i$- and $j$-particles ($\bm{r}_{ij}^{\rm
  (p)}$ in equation (\ref{eq:acc}), (\ref{eq:jrk}), and
(\ref{eq:pot})), and the accumulation of individual acceleration and
gravitational potential (summation in equation (\ref{eq:acc}) and
(\ref{eq:pot})) are computed in DP, and the remaining portions of the
force calculation are done in SP.

In computing accelerations, jerks, and potentials in equations
(\ref{eq:acc}), (\ref{eq:jrk}) and (\ref{eq:pot}), respectively, we
calculate the inverse square root of $(r_{ij}^{\rm (p)})^2+\epsilon^2$
using the VRSQRTPS instruction, which computes an approximated value
of the inverse square root very quickly to an accuracy of 12 bits. To
obtain an accuracy equivalent to SP, a Newton-Raphson iteration is
applied, such that
\begin{equation}
x_1 = - \frac{1}{2} x_0 (a x_0^2 - 3), \label{eq:newtonraphson}
\end{equation}
where $x_0$ is an initial guess for $1/\sqrt{a}$, and $x_1$ is
improved value of $1/\sqrt{a}$.  NMH06 reported that values returned
by the VRSQRTPS instruction contain statistical bias dependent on
implementation of this instruction. We statistically correct the bias
in the same way as NMH06.

\subsection{SIMD parallelization of the force calculation using the AVX instruction set}

Since we perform the force calculation mostly in SP, we calculate
eight interactions between $i$- and $j$-particles simultaneously.  In
our implementation, interactions between four $j$-particles and two
$i$-particles are calculated in parallel. Of course, there exist other
possible combinations of interactions such as four $i$-particles and
two $j$-particles. We have compared the performance of various
combinations of interactions and found that this choice exhibits the
best performance among others by $5$--$10$\%.


The calculation of forces on two $i$-particles (a force loop) in our
implementation consists of three parts as follows.
\begin{enumerate}
\item[1.] Prepare the $i$- and $j$-particle data in a suitable form
  for SIMD calculations.
\item[2.] Calculate the forces on the two $i$-particles exerted by
  all $j$-particles on the YMM registers.
\item[3.] Write back the calculated forces of the two $i$-particles
  on the YMM registers into the memory.
\end{enumerate}
The steps 1 and 3 are written in C language, while the step 2 is
written in C language with inline assembly language.

First, we describe the implementation of the steps 1 and 3. List
\ref{list:structures} shows the definitions of structures for $i$- and
$j$-particles as well as a structure for the results (accelerations,
jerks, and potentials) used in our implementation.  Before computing
the accelerations, jerks, and potentials of $i$-particles, the
positions, velocities, and indices of $i$-particles are stored in an
array of the structure, \verb|Ipart|, and the positions, velocities,
masses, and indices of $j$-particles are stored in an array of the
structure \verb|Jpart|. The calculated accelerations, jerks, and
potentials are stored into an array of the structure
\verb|NewAccJerk|.

\begin{lstlisting}[caption={Structures for $i$-particles, $j$-particles and the resulting accelerations.}, label={list:structures}]
struct Ipart{
  double xpos[8]; // (i0, i0, i0, i0, i1, i1, i1, i1)
  double ypos[8];
  double zpos[8];
  float  id[8];
  float  xvel[8];
  float  yvel[8];
  float  zvel[8];
};
struct Jpart{
  double xpos[4]; // (j0, j1, j2, j3)
  double ypos[4];
  double zpos[4];
  float  id[8];   // (j0, j1, j2, j3, j0, j1, j2, j3)
  float  mass[8];
  float  xvel[8];
  float  yvel[8];
  float  zvel[8];
};
struct NewAccJerk{
  double xacc;
  double yacc;
  double zacc;
  double pot;
  float  xjrk;
  float  yjrk;
  float  zjrk;
};
\end{lstlisting}

\begin{figure}
  \begin{center}
    \includegraphics[scale=0.5]{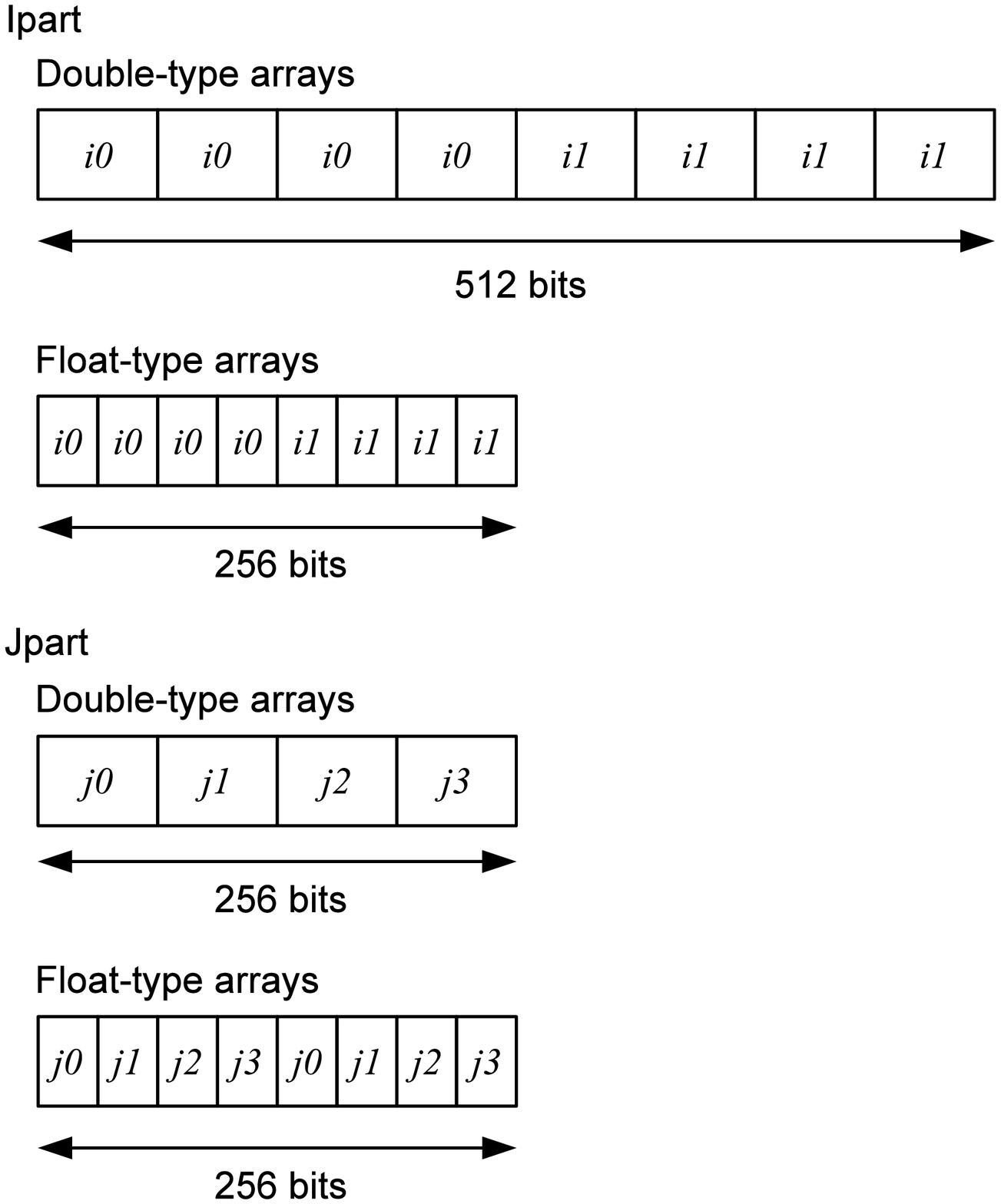}
  \end{center}
  \caption{Assignments of the particle data in the structures of $i$-
  and $j$-particles.}  \label{fig:array}
\end{figure}

Note that the size of each array in the structures of $i$- and
$j$-particles is multiple of $256$ bits, the length of the YMM
registers, so that each array can be readily loaded onto the YMM
registers. The structures of $i$-particles (\verb|Ipart|) and
$j$-particles (\verb|Jpart|) contain the data of two $i$-particles and
four $j$-particles, respectively. In figure \ref{fig:array}, we show
the assignments of particle indices in the arrays in these structures,
where $i0$ and $i1$, and $j0$, $j1$, $j2$, and $j3$ are the indices of
$i$- and $j$-particles, respectively. Such assignments enable us to
calculate the accelerations, jerks, and potentials of two
$i$-particles from four $j$-particles in a efficient manner with the
AVX instructions. In the step 1, we arrange the particle data and
substitute them into the structures \verb|Ipart| and \verb|Jpart|
before calculating the accelerations and jerks in the step 2.

Next, we describe the implementation of the step 2, in which the
accelerations, jerks, and potentials of $i$-particles are computed
using the AVX instructions. In issuing the AVX instructions, we use
the inline assembly language and manually control the assignment of
the YMM registers to the computational data to achieve a good
performance. We also try to obtain a high issue rate of the AVX
instructions by optimizing the order of operations so that operands in
adjacent instruction calls do not have dependencies as much as
possible.

For the readability of the source code, we introduce a set of
preprocessor macros of the AVX instructions. The macros used in our
implementation are shown in List \ref{list:macros}. Most of them have
three operands, in which \verb|src1| and \verb|src2| are the source
operands and the results of the instructions are stored in the last
operand (destination operand, \verb|dst|). In these macros, operands
named \verb|src|, \verb|src1|, \verb|src2| and \verb|dst| are
in the XMM/YMM registers, and those named \verb|mem| indicate the main
memory or the cache memory. Brief descriptions of these macros are
summarized in Table \ref{tab:description}. More detailed documents on
the AVX instructions can be found in Intel's
website.\footnote{http://software.intel.com/en-us/avx/}

\begin{lstlisting}[caption={Preprocessor macros for inline assembly codes.}, label={list:macros}]
#define VZEROALL asm("vzeroall");
#define VLOADPS(mem, dst) \
  asm("vmovaps %0, %"dst::"m"(mem));
#define VLOADPD(mem, dst) \
  asm("vmovapd %0, %"dst::"m"(mem));
#define VSTORPS(src, mem) \
  asm("vmovaps %"src ", %0" ::"m"(mem));
#define VSTORPD(src, mem) \
  asm("vmovapd %"src ", %0" ::"m"(mem));
#define VADDPS(src1, src2, dst) \
  asm("vaddps " src1 "," src2 "," dst);
#define VADDPD(src1, src2, dst) \
  asm("vaddpd " src1 "," src2 "," dst);
#define VSUBPS(src1, src2, dst) \
  asm("vsubps " src1 "," src2 "," dst);
#define VSUBPS_M(mem, src, dst) \
  asm("vsubps %0, %"src ", %"dst " "::"m"(mem));
#define VSUBPD(src1, src2, dst) \
  asm("vsubpd " src1 "," src2 "," dst);
#define VSUBPD_M(mem, src, dst) \
  asm("vsubpd %0, %"src ", %"dst " "::"m"(mem));
#define VMULPS(src1, src2, dst) \
  asm("vmulps " src1 "," src2 "," dst);
#define VMULPS_M(mem, src, dst) \
  asm("vmulps %0, %"src ", %"dst " "::"m"(mem));
#define VHADDPD(src1, src2, dst) \
  asm("vhaddpd " src1 "," src2 "," dst);
#define VRSQRTPS(src, dst) \
  asm("vrsqrtps " src "," dst);
#define VCVTPD2PS(src, dst) \
  asm("vcvtpd2ps " src "," dst);
#define VCVTPS2PD(src, dst) \
  asm("vcvtps2pd " src "," dst);
#define VMERGE(src1, src2, dst) \
  asm("vperm2f128 %0, %"src1 ", %"src2 ", \
%"dst " "::"g"(2));
#define VUP2LOW(src, dst) \
  asm("vextractf128 %0, %"src ", %"dst " "::"g"(1));
#define VANDPS(src1, src2, dst) \
  asm("vandps " src1 "," src2 "," dst);
#define VCMPNEQPS(src1, src2, dst) \
  asm("vcmpps %0, %"src1 ", %"src2 ", %"dst " \
"::"g"(4));
#define PREFETCH(mem) asm("prefetcht0 %0"::"m"(mem))
\end{lstlisting}

\begin{table*}
  \begin{center}
   \caption{Descriptions of the macros for inline assembly codes.
     ``src'', ``src1'', ``src2'' and ``dst'' indicate the YMM
     registers and ``mem'' indicates the main memory or the cache
     memory.}
    \begin{tabular}{|l|l|}
     \hline
     ~\verb|VZEROALL| & ~zero out all registers \\ \hline
     ~\verb|VLOADPS(mem,dst)| & ~load 8 SP data from \verb|mem| into \verb|dst| \\ \hline 
     ~\verb|VLOADPD(mem,dst)| & ~load 4 DP data from \verb|mem| into \verb|dst| \\ \hline
     ~\verb|VSTORPS(src,mem)| & ~store 8 SP data in \verb|src| into \verb|mem| \\ \hline 
     ~\verb|VSTORPD(src,mem)| & ~store 4 DP data in \verb|src| into \verb|mem| \\ \hline
     ~\verb|VADDPS(src1,src2,dst)| & ~add 8 SP data in \verb|src1| and \verb|src2| and store the results to \verb|dst| \\ \hline
     ~\verb|VADDPD(src1,src2,dst)| & ~add 4 DP data in \verb|src1| and \verb|src2| and store the results to \verb|dst| \\ \hline
     ~\verb|VSUBPS(src1,src2,dst)| & ~subtract 8 SP data in \verb|src1| from 8 SP data in \verb|src2| and store the results to \verb|dst| \\ \hline 
     ~\verb|VSUBPS_M(mem,src,dst)| & ~subtract 8 SP data in \verb|mem| from 8 SP data in \verb|src| and store the results to \verb|dst| \\ \hline 
     ~\verb|VSUBPD_M(mem,src,dst)| & ~subtract 4 DP data in \verb|mem| from 4 DP data in \verb|src| and store the results to \verb|dst| \\ \hline 
     ~\verb|VMULPS(src1,src2,dst)| & ~multiply 8 SP data in \verb|src1| and \verb|src2| and store the results to \verb|dst| \\ \hline 
     ~\verb|VMULPS_M(mem,src,dst)| & ~multiply 8 SP data in \verb|mem| and \verb|src| and store the results to \verb|dst| \\ \hline 
     ~\verb|VHADDPD(src1,src2,dst)| & ~add pairs of adjacent SP data in each of \verb|src1| and \verb|src2| and store the results to \verb|dst| \\ \hline 
     ~\verb|VRSQRTPS(src,dst)| & ~return the inverse square root of 8 SP data in \verb|src| and store the results to \verb|dst| \\ \hline 
     ~\verb|VCVTPD2PS(src,dst)| & ~convert 4 DP data in \verb|src| to 4 SP data and store the results to the lower 128 bits of \verb|dst| \\ \hline 
     ~\verb|VCVTPS2PD(src,dst)| & ~convert 4 SP data in lower 128 bits of \verb|src| to 4 DP data and store the results to \verb|dst| \\ \hline 
     ~\verb|VMERGE(src1,src2,dst)| & ~concatenate data in the lower 128 bits in \verb|src1| and \verb|src2| and store the results to \verb|dst| \\ \hline 
     ~\verb|VUP2LOW(src,dst)| & ~copy the upper 128 bits in \verb|src| to the lower 128 bits in \verb|dst| \\ \hline 
     ~\verb|VANDPS(src1,src2,dst)| & ~operate bitwise logical AND on 8 SP data in \verb|src1| and \verb|src2| and store the results to \verb|dst| \\ \hline 
     ~\verb|VCMPNEQPS(src1,src2,dst)| & ~compare 8 SP data in \verb|src1| and \verb|src2| and  the unequal field of \verb|dst| \\ \hline 
     ~\verb|PREFETCH(mem)| & ~prefetch data on \verb|mem| to the cache memory by one cache line (64 bytes)\\ \hline
    \end{tabular}
    \label{tab:description}
  \end{center}
\end{table*}

Table \ref{tab:register} shows the assignment of the YMM registers to
the physical quantities in the calculations of accelerations, jerks,
and potentials, where the subscripts of $x$, $y$ and $z$ indicates
$x$, $y$, and $z$ components of vectors, respectively. Hereafter, for
simplicity, we omit the superscripts (p) and (1) of $\bm{r}$ and
$\bm{v}$, and of $\bm{a}$, $\bm{j}$ and $\phi$, respectively. With
this assignment of the YMM registers, we can carry out the computation
using the data in only the YMM registers except for one load
instruction for each $j$-particle data.

\begin{table}
  \begin{center}
    \caption{Aliases of the YMM registers and the assignment of
      variables to the registers.}
    \begin{tabular}{|c|c|c|}
      \hline Alias & ID & Variables \\ \hline\verb|YMM00| & \verb|%ymm0| &
      $\bm{r}_{ij} \cdot \bm{v}_{ij}$, $\bm{ r}_{ij} \cdot
      \bm{v}_{ij}/\tilde{r}_{ij}^2$ \\ \hline \verb|YMM01| & \verb|%ymm1| &
      $\tilde{r}_{ij}^2$, $1/\tilde{r}_{ij}$ \\ \hline \verb|YMM02| & \verb|%ymm2| &
      $Gm_j / \tilde{r}_{ij}$, $Gm_j / \tilde{r}_{ij}^3$ \\ \hline \verb|YMM03| &
      \verb|%ymm3| & $r_{ij,x}$ \\ \hline \verb|YMM04| &
      \verb|%ymm4| & $r_{ij,y}$ \\ \hline \verb|YMM05| &
      \verb|%ymm5| & $r_{ij,z}$ \\ \hline \verb|YMM06| &
      \verb|%ymm6| & $v_{ij,x}$ \\ \hline \verb|YMM07| &
      \verb|%ymm7| & $v_{ij,y}$ \\ \hline \verb|YMM08| &
      \verb|%ymm8| & $v_{ij,z}$ \\ \hline \verb|YMM09| &
      \verb|%ymm9| & $\phi_i$\\ \hline \verb|YMM10| & \verb|%ymm10| &
      $a_{i,x}$ \\ \hline \verb|YMM11| & \verb|%ymm11| &
      $a_{i,y}$ \\ \hline \verb|YMM12| & \verb|%ymm12| &
      $a_{i,z}$ \\ \hline \verb|YMM13| & \verb|%ymm13| &
      $j_{i,x}$ \\ \hline \verb|YMM14| & \verb|%ymm14| &
      $j_{i,y}$ \\ \hline \verb|YMM15| & \verb|%ymm15| &
      $j_{i,z}$ \\ \hline
    \end{tabular}
    \label{tab:register}
  \end{center}
\end{table}

\begin{figure*}
 \begin{center}
  \includegraphics[scale=0.8]{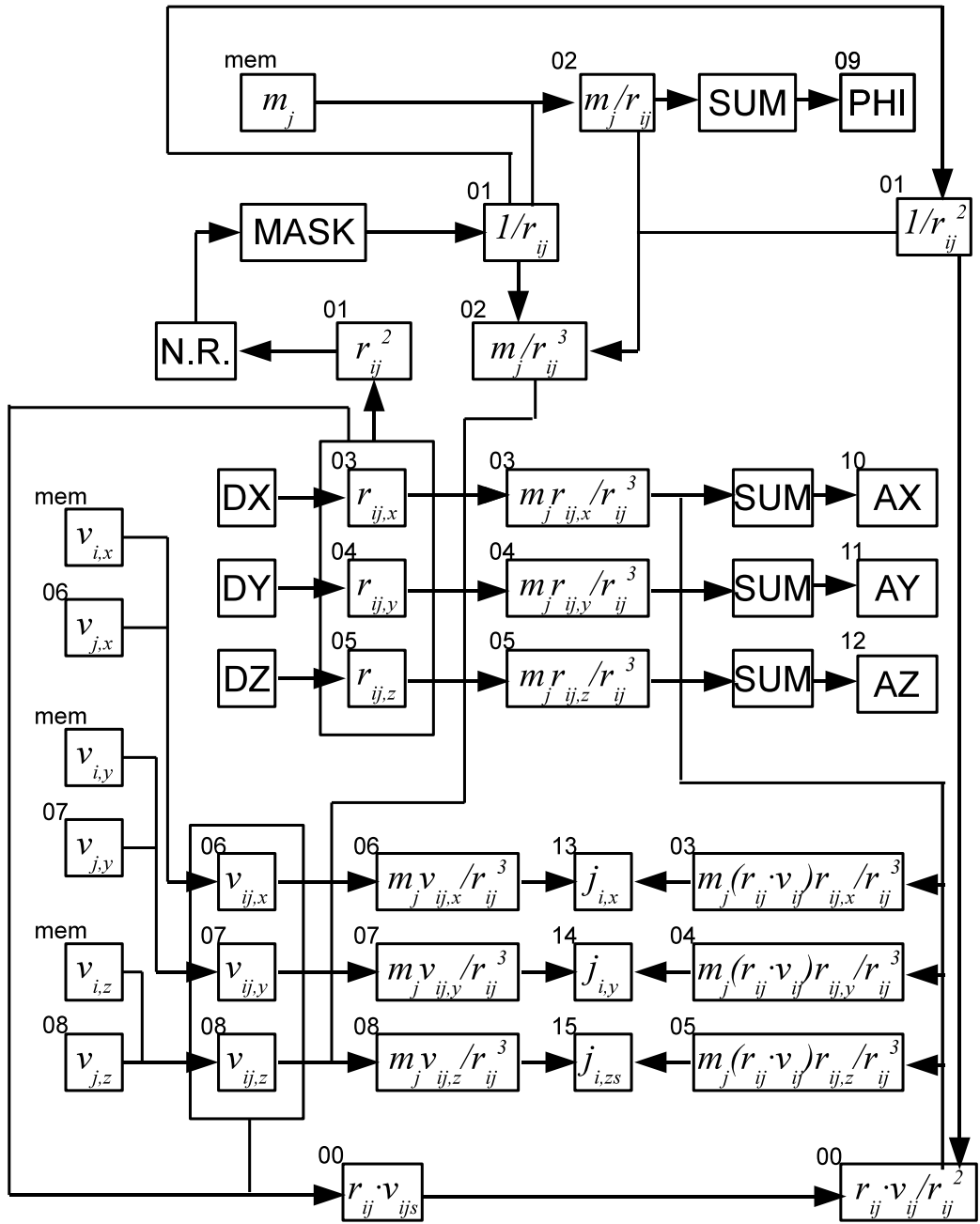}
 \end{center}
 \caption{A schematic illustration of the force loop. Numbers at upper
   left of each box indicate indices of the YMM registers in which data
   are stored, and "mem" indicates the main memory. Data assignment in
   all the registers are the same, and are drawn in figure
   \ref{fig:datainymmregister}.}
 \label{fig:schematic}
\end{figure*}

\begin{figure}
 \begin{center}
  \includegraphics[scale=0.5]{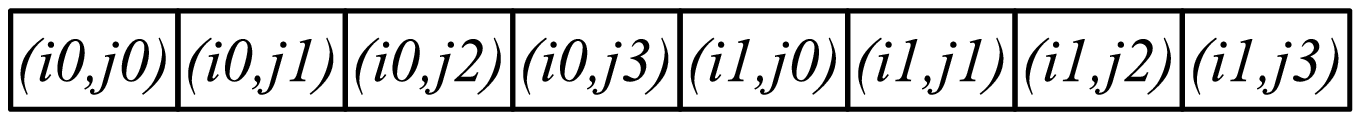}
 \end{center}
 \caption{Data assignment in the YMM registers in figure \ref{fig:schematic}.}
 \label{fig:datainymmregister}
\end{figure}

Figure \ref{fig:schematic} shows a schematic illustration of a force
loop to calculate accelerations, jerks, and potentials on two
$i$-particles from four $j$-particles by using the AVX
instructions. The variables with a subscript $i$ and $j$ in
figure~\ref{fig:schematic} contains the $8$ SP data of $i$- and
$j$-particles in the same order of indices shown in
figure~\ref{fig:array}, and those with a subscript $ij$ in
figure~\ref{fig:schematic} such as $r_{ij}$ and $v_{ij}$ contains the
data with indices of $i$- and $j$-particles in the order shown in
figure~\ref{fig:datainymmregister}. In figure~\ref{fig:schematic}, the
blocks \verb|DX|, \verb|DY| and \verb|DZ| indicate the calculation of
relative coordinates between two $i$-particles and four $j$-particles
for $x$, $y$ and $z$ components, respectively, in the mixed precision,
and the block \verb|SUM| indicates the accumulation of physical
quantities (accelerations and potentials) of two $i$-particles
interacting with four $j$-particles in the DP. The details of the
blocks \verb|DX|, \verb|DY|, \verb|DZ| and \verb|SUM| are described
later.

First we describe the procedure depicted in figure
\ref{fig:schematic}.

\begin{figure*}
  \begin{center}
    \includegraphics[scale=0.8]{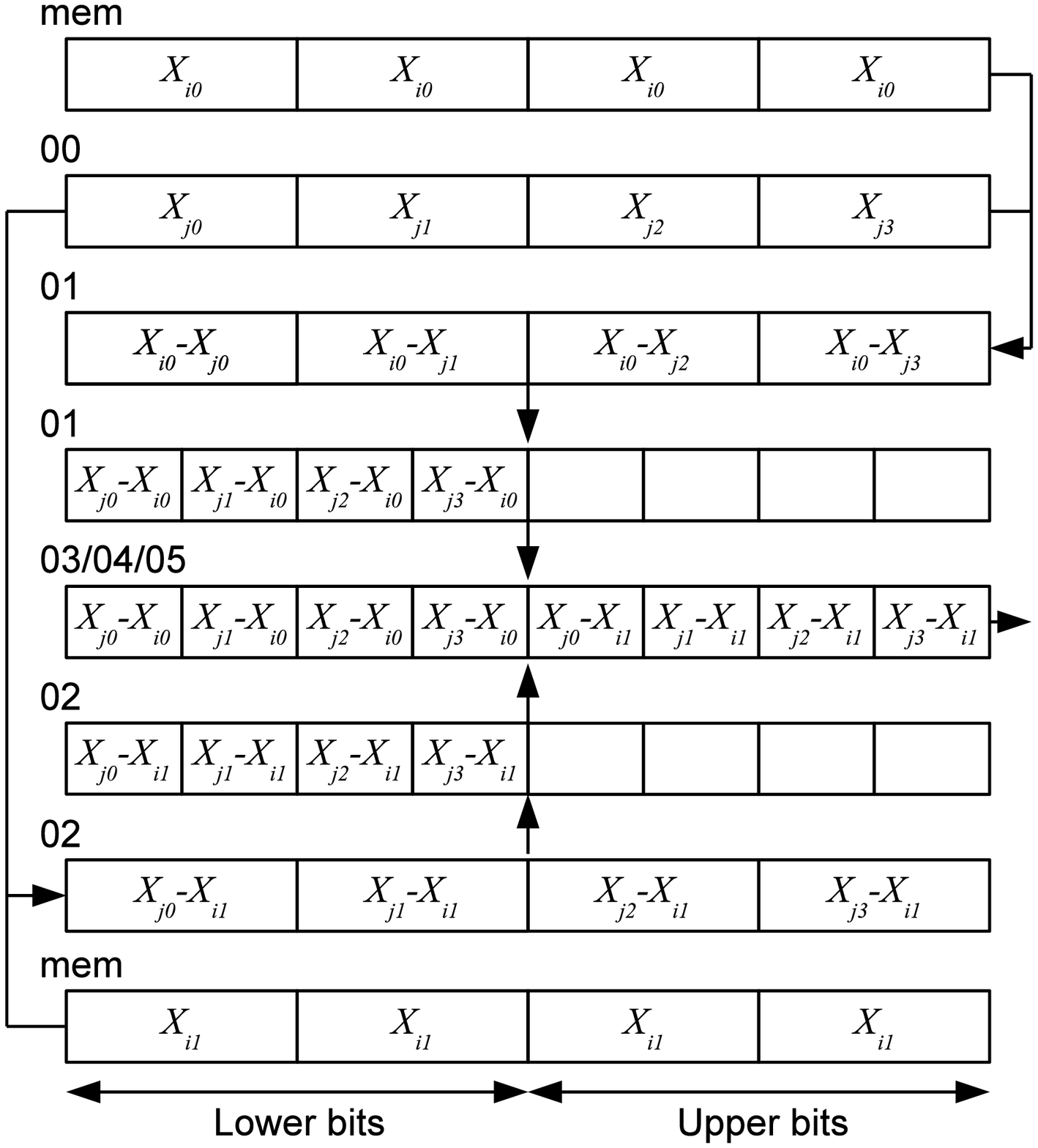}
  \end{center}
  \caption{A schematic illustration of the blocks DX, DY, and DZ
    displayed in figure \ref{fig:schematic}. Numbers at upper left of
    each box indicate indices of the YMM registers in which data are
    stored, and "mem" indicates the main memory.}  \label{fig:dx}
\end{figure*}

\begin{figure*}
  \begin{center}
    \includegraphics[scale=0.8]{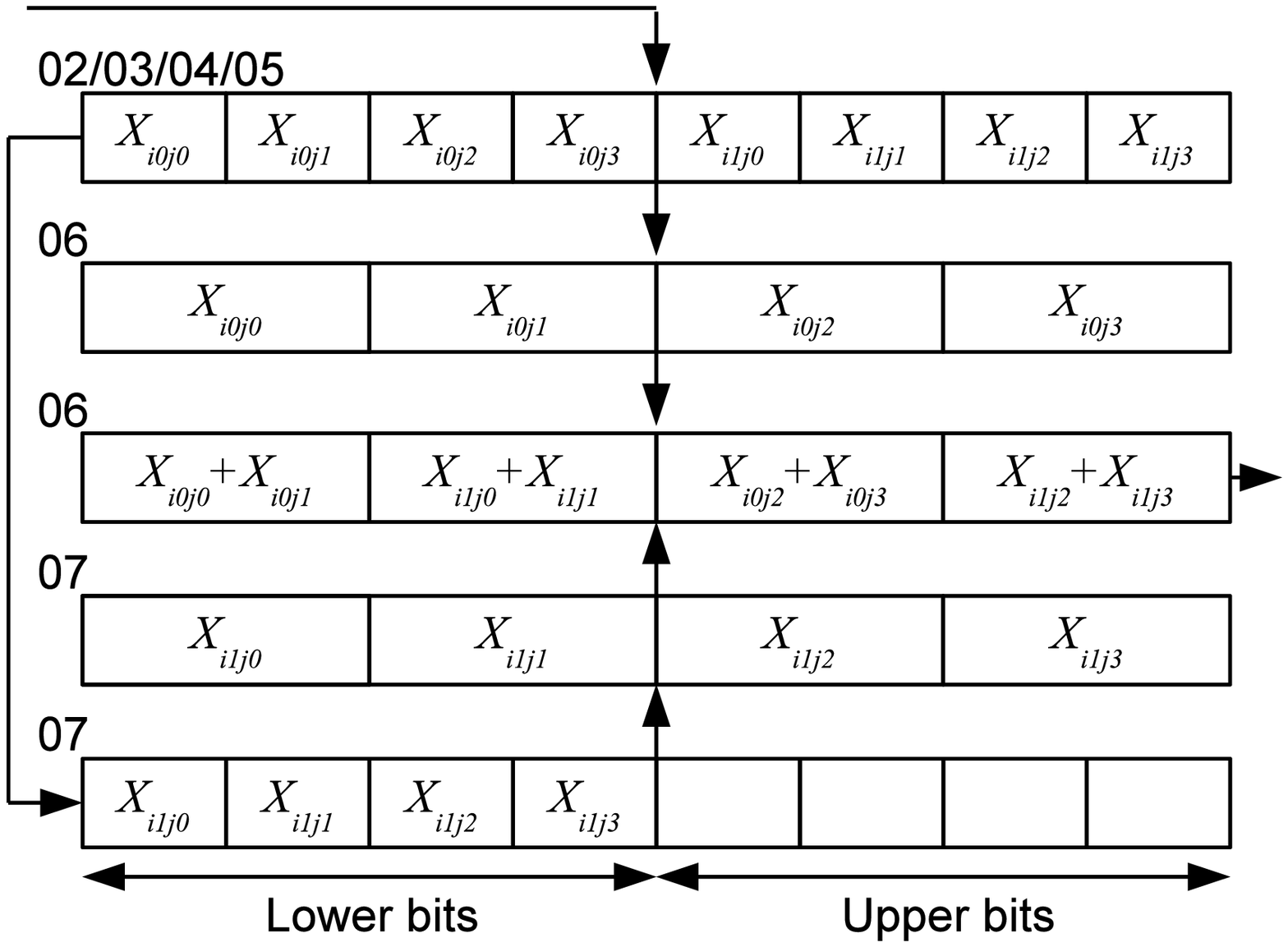}
  \end{center}
  \caption{A schematic illustration of SUM shown in figure
    \ref{fig:schematic}. Numbers at upper left of each box indicate
    indices of the YMM registers in which data are stored, and "mem"
    indicates the main memory.}  \label{fig:sum}
\end{figure*}

\begin{enumerate}
 \item[1.] Calculate $r_{ij,x}$, $r_{ij,y}$, and $r_{ij,z}$ in
   \verb|DX|, \verb|DY|, and \verb|DZ| (see figure~\ref{fig:dx}), and
   store them into \verb|YMM03|, \verb|YMM04|, and \verb|YMM05|,
   respectively.
 \item[2.] Square $r_{ij,x}$ in \verb|YMM03|, $r_{ij,y}$ in
   \verb|YMM04|, and $r_{ij,z}$ in \verb|YMM05|, and then sum up them
   and the squared softening length to compute the softened squared
   distance, $\tilde{r}_{ij}^2\equiv r_{ij}^2+\epsilon^2$, and store
   it into \verb|YMM01|.
 \item[3.] Calculate an inverse square root of $\tilde{r}^2_{ij}$,
   operate Newton-Raphson iteration for that value (the block
   \verb|N.R.| in figure~\ref{fig:schematic}), and store the result
   $1/\tilde{r}_{ij}$ into \verb|YMM01|. If indices of $i$- and $j$-
   particles are the same, zero is stored in the corresponding element
   of \verb|YMM01| (the block \verb|MASK| in
   figure~\ref{fig:schematic}).
 \item[4.] Multiply $1/\tilde{r}_{ij}$ in \verb|YMM01| by $m_j$ in the
   main memory obtaining the gravitational potential exerted by
   $j$-particles $\phi_{ij}\equiv m_j/\tilde{r}_{ij}$, and store the
   result into \verb|YMM02|.
 \item[5.] Calculate the summation of $m_j/\tilde{r}_{ij}$ in
   \verb|YMM02| over the subscript $j$ (the block \verb|SUM| in figure
   \ref{fig:schematic}), and accumulate the results into \verb|YMM09|
   (the block \verb|PHI| in figure~\ref{fig:schematic}).  (see also
   figure~\ref{fig:sum})
 \item[6.] Load $v_{j,x}$, $v_{j,y}$, and $v_{j,z}$ to \verb|YMM06|,
   \verb|YMM07|, and \verb|YMM08|, respectively.
 \item[7.] Subtract $v_{i,x}$, $v_{i,y}$, and $v_{i,z}$ in the main
   memory from $v_{j,x}$ in \verb|YMM06|, $v_{j,y}$ in \verb|YMM07|,
   and $v_{j,z}$ in \verb|YMM08|, and store the results ($v_{ij,x}$,
   $v_{ij,y}$, and $v_{ij,z}$) into \verb|YMM06|, \verb|YMM07|, and
     \verb|YMM08|, respectively.
 \item[8.] Calculate inner product of relative position and relative
   velocity vectors, using $r_{ij,x}$ in \verb|YMM03|,
   $r_{ij,y}$ in \verb|YMM04|, $r_{ij,z}$ in \verb|YMM05|,
   $v_{ij,x}$ in \verb|YMM06|, $v_{ij,y}$ in \verb|YMM07|,
   and $v_{ij,z}$ in \verb|YMM08|, and store the result
   ($\bm{r}_{ij} \cdot \bm{v}_{ij}$) into \verb|YMM00|.
 \item[9.] Square $1/\tilde{r}_{ij}$ in \verb|YMM01|, and
	   store the result $1/\tilde{r}_{ij}^2$ into \verb|YMM01|.
 \item[10.] Multiply $m_j/\tilde{r}_{ij}$ in \verb|YMM02| and
   $\bm{r}_{ij} \cdot \bm{v}_{ij}$ in \verb|YMM00| by
   $1/\tilde{r}_{ij}^2$ in \verb|YMM01|, and store the results
   ($m_j/\tilde{r}_{ij}^3$ and $\bm{ r}_{ij} \cdot
   \bm{v}_{ij}/\tilde{r}_{ij}^2$) into \verb|YMM02| and \verb|YMM00|,
   respectively.
 \item[11.] Multiply $r_{ij,x}$ in \verb|YMM03|, $r_{ij,y}$
   in \verb|YMM04|, and $r_{ij,z}$ in \verb|YMM05| by
   $m_j/\tilde{r}_{ij}^3$ in \verb|YMM02| obtaining the acceleration
   vector $\bm{a}_{ij}$ exerted by $j$-particles, and store the
   results into \verb|YMM03|, \verb|YMM04|, and \verb|YMM05|,
   respectively.
 \item[12.] Calculate the summation of
   $m_jr_{ij,x}/\tilde{r}_{ij}^3$ in \verb|YMM03|,
   $m_jr_{ij,y}/\tilde{r}_{ij}^3$ in \verb|YMM04|, and
   $m_jr_{ij,z}/\tilde{r}_{ij}^3$ in \verb|YMM05| over the
   subscript $j$ (the block \verb|SUM| in figure \ref{fig:schematic}),
   and accumulate the results into in \verb|YMM10|, \verb|YMM11|, and
   \verb|YMM12|, respectively. These correspond to the blocks
   \verb|AX|, \verb|AY|, and \verb|AZ| in figure \ref{fig:schematic},
   respectively.
 \item[13.] Multiply $v_{ij,x}$ in \verb|YMM06|, $v_{ij,y}$
   in \verb|YMM07|, and $v_{ij,z}$ in \verb|YMM08| by
   $m_j/\tilde{r}_{ij}^3$ in \verb|YMM02|, obtaining the first term of
   the jerks in equation (\ref{eq:jrk}), and store the results into
   \verb|YMM06|, \verb|YMM07|, and \verb|YMM08|, respectively.
 \item[14.] Accumulate the first term of the jerks
   $m_jv_{ij,x}/\tilde{r}_{ij}^3$ in \verb|YMM06|,
   $m_jv_{ij,y}/\tilde{r}_{ij}^3$ in \verb|YMM07|, and
   $m_jv_{ij,z}/\tilde{r}_{ij}^3$ in \verb|YMM08| into
   \verb|YMM13|, \verb|YMM14|, and \verb|YMM15|, respectively.
 \item[15.] Multiply $m_jr_{ij,x}/\tilde{r}_{ij}^3$ in
   \verb|YMM03|, $m_jr_{ij,y}/\tilde{r}_{ij}^3$ in \verb|YMM04|,
   and $m_jr_{ij,z}/\tilde{r}_{ij}^3$ in \verb|YMM05| by
   $\bm{r}_{ij} \cdot \bm{v}_{ij}/\tilde{r}_{ij}^2$ in \verb|YMM00|
   obtaining the second term of jerks in equation (\ref{eq:jrk}), and
   store the results into \verb|YMM03|, \verb|YMM04|, and
   \verb|YMM05|, respectively.
 \item[16.] Accumulate the second term of jerks
   $m_j(\bm{r}_{ij}\cdot\bm{ v}_{ij})r_{ij,x}/\tilde{r}_{ij}^5$
   in \verb|YMM03|, $m_j(\bm{r}_{ij}
   \cdot\bm{v}_{ij})r_{ij,y}/\tilde{r}_{ij}^5$ in \verb|YMM04|,
   and
   $m_j(\bm{r}_{ij}\cdot\bm{v}_{ij})r_{ij,z}/\tilde{r}_{ij}^5$ in
   \verb|YMM05| into \verb|YMM13|, \verb|YMM14|, and \verb|YMM15|,
   respectively.
 \item[17.] Return to the step 1 until all the $j$-particles are
   processed.
\end{enumerate}

\vspace{0.3cm} 

The calculation of the relative coordinates between two $i$-particles
and four $j$-particles done in the blocks \verb|DX|, \verb|DY|, and
\verb|DZ| are depicted in figure~\ref{fig:dx}.  As an example, the
calculation of $r_{ij,x}$ are processed as follows.
\begin{enumerate}
 \item[1.] Load four $r_{j,x}$ ($j=j0 \cdots j3$) to \verb|YMM00|
   from the main memory in DP.
 \item[2.] Subtract the coordinate of the first $i$-particles
   $r_{i0,x}$ in the main memory from $r_{j,x}$ in
   \verb|YMM00|, and store the result, $r_{i0j,x}$, in
   \verb|YMM01|.
 \item[3.] Subtract the coordinate of the second $i$-particle
   $r_{i1,x}$ in the main memory from $r_{j,x}$ in
   \verb|YMM00|, and store the result, $r_{i1j,x}$, in
   \verb|YMM02|.
 \item[4.] Convert type of $r_{i0j,x}$ in \verb|YMM01| from DP to
   SP, and store the result in the lower bits of \verb|YMM01|.
 \item[5.] Convert type of $r_{i1j,x}$ in \verb|YMM02| from DP to
   SP, and store the result in the lower bits of \verb|YMM02|.
 \item[6.] Copy $r_{i0j,x}$ in the lower bits of \verb|YMM01| to
   the lower bits of \verb|YMM03|, and $r_{i1j,x}$ in the lower
   bits of \verb|YMM02| to the upper bits of \verb|YMM03|.
\end{enumerate}

\vspace{0.3cm} 

Figure~\ref{fig:sum} illustrates the procedure to accumulate the
accelerations and potentials on two $i$-particles exerted by four
$j$-particles in DP. In the step 2, eight accelerations $\bm{a}_{ij}$
and potentials $\phi_{ij}$ between two $i$-particles and four
$j$-particles are calculated on the YMM registers in SP. Note that jerks
are accumulated in SP, instead of DP. In fact, we cannot compute the
summations of accelerations and potentials over all four $j$-particles
in DP using the AVX instructions. Instead, two partial summations over
two $j$-particles can be computed for each $i$-particle.  In
accumulating gravitational potentials, for example, the procedures are
given as follows.
\begin{enumerate}
 \item[1.] Convert $\phi_{i0j}$ ($j=j0\cdots j3$) in the lower 128
   bits of \verb|YMM02| from SP to DP, and store the result in
   \verb|YMM06|.
 \item[2.] Copy $\phi_{i1j}$ ($j=j0\cdots j3$) in the upper 128 bits
   of \verb|YMM02| to the lower 128 bits of \verb|YMM07|.
 \item[3.] Convert $\phi_{i1j}$ in the lower 128 bits of \verb|YMM07|
   from SP to DP, and store the result in \verb|YMM07|.
 \item[4.] Operate a horizontal sum reduction of \verb|YMM06| and
   \verb|YMM07|, obtaining $\phi_{i0j0}+\phi_{i0j1}$,
   $\phi_{i1j0}+\phi_{i1j1}$, $\phi_{i0j2}+\phi_{i0j3}$, and
   $\phi_{i1j2}+\phi_{i1j3}$ and store them in \verb|YMM06|.
 \item[5.] Accumulate the partially reduced potentials of
   $i$-particles with indices of $i0$ and $i1$ in \verb|YMM06| to
   $\phi_i$ in \verb|YMM09|.
\end{enumerate}

\vspace{0.3cm}

The function \verb|calc_accjerkpot| which computes the forces on two
$i$-particles from all the $j$-particles is shown in List
\ref{list:newtonforce}. In the force loop at line 15 in List
\ref{list:newtonforce}, the accelerations, jerks, and potentials on
$i$-particles are calculated as we described.  Before entering the
force loop, $i$-particles are set, the pointer to $j$-particles is set
to point the first element of the array of \verb|Jpart|, and all the
registers are set to zero.  Once the force loop has been finished, sum
reductions are operated on the partially accumulated accelerations,
jerks, and potentials, and the results are stored in components of the
array of structure \verb|NewAccJerk|. The resulting accelerations,
jerks, and potentials are multiplied by some coefficients. This is
because we omit the multiplication of $-1/2$ in equation
(\ref{eq:newtonraphson}) in the force loop. (In this source code, we
do not correct the bias in the VRSQRTPS instruction for simplicity.)

As seen in List \ref{list:newtonforce}, some operands are specified by
\verb|XMM|, which are aliases of the XMM registers. In some instructions,
the lower bits of the YMM register is only used for their operands. When
such instructions are operated, the XMM registers have to be specified.

\begin{lstlisting}[caption={A force loop to calculate the Newton's force using the AVX instructions}, label={list:newtonforce}]
void calc_accjerkpot(struct Ipart *ipart,
                     struct Jpart *jpart,
                     struct NewAccJerk *accjerk)
{
  double xacc[4], yacc[4], zacc[4], pot[4];
  float  xjrk[8], yjrk[8], zjrk[8];
  float veps2[8] = {eps2, eps2, ... , eps2};
  float three[8] = {3.0, 3.0, ... , 3.0};
  float threefourth[8] = {0.75, 0.75, ... , 0.75};
  struct Ipart *iptr = ipart;
  struct Jpart *jptr = jpart;

  // initialize
  VZEROALL;
  for(j = 0; j < nj; j += 4, jptr++){
    // r_ij,x -> YMM03
    VLOADPD(jptr->xpos[0], YMM00);
    VSUBPD_M(ipart->xpos[0], YMM00, YMM01);
    VSUBPD_M(iptr->xpos[4], YMM00, YMM02);
    VCVTPD2PS(YMM01, XMM01);
    VCVTPD2PS(YMM02, XMM02);
    VMERGE(YMM01, YMM02, YMM03);
    // r_ij,y -> YMM04
    VLOADPD(jptr->ypos[0], YMM00);
    VSUBPD_M(iptr->ypos[0], YMM00, YMM01);
    VSUBPD_M(iptri->ypos[4], YMM00, YMM02);
    VCVTPD2PS(YMM01, XMM01);
    VCVTPD2PS(YMM02, XMM02);
    VMERGE(YMM01, YMM02, YMM04);
    // r_ij,z -> YMM05
    VLOADPD(jptr->zpos[0], YMM00);
    VSUBPD_M(iptr->zpos[0], YMM00, YMM01);
    VSUBPD_M(iptr->zpos[4], YMM00, YMM02);
    VCVTPD2PS(YMM01, XMM01);
    VCVTPD2PS(YMM02, XMM02);
    VMERGE(YMM01, YMM02, YMM05);
    // (r_ij)^2 -> YMM01
    VLOADPS(veps2[0], YMM01);
    VMULPS(YMM03, YMM03, YMM00);
    VADDPS(YMM00, YMM01, YMM01);
    VMULPS(YMM04, YMM04, YMM00);
    VADDPS(YMM00, YMM01, YMM01);
    VMULPS(YMM05, YMM05, YMM00);
    VADDPS(YMM00, YMM01, YMM01);
    // - 2 / r_ij -> YMM01
    VRSQRTPS(YMM01, YMM02);
    VMULPS(YMM02, YMM01, YMM01);
    VMULPS(YMM02, YMM01, YMM01);
    VSUBPS_M(three[0], YMM01, YMM01);
    VMULPS(YMM02, YMM01, YMM01);
    // exclude self interaction
    VLOADPS(jptr->id[0], YMM02);
    VLOADPS(iptr->id[0], YMM00);
    VCMPNEQPS(YMM00, YMM02, YMM02);
    VANDPS(YMM02, YMM01, YMM01);
    // phi_i -> YMM09
    VMULPS_M(jptr->mass[0], YMM01, YMM02);
    VCVTPS2PD(XMM02, YMM00);
    VUP2LOW(YMM02, XMM06);
    VCVTPS2PD(XMM06, YMM06);
    VHADDPD(YMM06, YMM00, YMM07);
    VADDPD(YMM07, YMM09, YMM09);
    // v_ij,x, v_ij,y, v_ij,z
    VLOADPS(jptr->xvel[0], YMM06);
    VSUBPS_M(iptr->xvel[0], YMM06, YMM06);
    VLOADPS(jptr->yvel[0], YMM07);
    VSUBPS_M(iptr->yvel[0], YMM07, YMM07);
    VLOADPS(jptr->zvel[0], YMM08);
    VSUBPS_M(iptr->zvel[0], YMM08, YMM08);
    // r_ij * v_ij -> YMM00
    VMULPS(YMM03, YMM06, YMM00);
    VMULPS(YMM04, YMM07, YMM02);
    VADDPS(YMM02, YMM00, YMM00);
    VMULPS(YMM05, YMM08, YMM02);
    VADDPS(YMM02, YMM00, YMM00);
    // 3.0 * r_ij * v_ij / (r_ij)^2 -> YMM00
    // - m_j / r_ij^3 -> YMM02
    VMULPS_M(jptr->mass[0], YMM01, YMM02);
    VMULPS(YMM01, YMM01, YMM01);
    VMULPS(YMM01, YMM00, YMM00);
    VMULPS(YMM01, YMM02, YMM02);
    VMULPS_M(threefourth[0], YMM00, YMM00);
    // prefetch
    PREFETCH((jptr+1)->xpos[0]);
    PREFETCH((jptr+1)->zpos[0]);
    PREFETCH((jptr+1)->mass[0]);
    PREFETCH((jptr+1)->yvel[0]);
    // j_i,x += j_ij,x (first term)
    VMULPS(YMM02, YMM06, YMM06);
    VADDPS(YMM06, YMM13, YMM13);
    // j_i,y += j_ij,y (first term)
    VMULPS(YMM02, YMM07, YMM07);
    VADDPS(YMM07, YMM14, YMM14);
    // j_i,z += j_ij,z (first term)
    VMULPS(YMM02, YMM08, YMM08);
    VADDPS(YMM08, YMM15, YMM15);
    // ax_i,x += ax_ij,x
    VMULPS(YMM02, YMM03, YMM03);
    VCVTPS2PD(XMM03, YMM06);
    VUP2LOW(YMM03, XMM07);
    VCVTPS2PD(XMM07, YMM07);
    VHADDPD(YMM07, YMM06, YMM06);
    VADDPD(YMM06, YMM10, YMM10);
    // ay_i,y += ay_ij,y
    VMULPS(YMM02, YMM04, YMM04);
    VCVTPS2PD(XMM04, YMM06);
    VUP2LOW(YMM04, XMM07);
    VCVTPS2PD(XMM07, YMM07);
    VHADDPD(YMM07, YMM06, YMM06);
    VADDPD(YMM06, YMM11, YMM11);
    // az_i,z += az_ij,z
    VMULPS(YMM02, YMM05, YMM05);
    VCVTPS2PD(XMM05, YMM06);
    VUP2LOW(YMM05, XMM07);
    VCVTPS2PD(XMM07, YMM07);
    VHADDPD(YMM07, YMM06, YMM06);
    VADDPD(YMM06, YMM12, YMM12);
    // j_i,x += j_ij,x (second term)
    VMULPS(YMM00, YMM03, YMM03);
    VSUBPS(YMM03, YMM13, YMM13);
    // j_i,y += j_ij,y (second term)
    VMULPS(YMM00, YMM04, YMM04);
    VSUBPS(YMM04, YMM14, YMM14);
    // j_i,z += j_ij,z (second term)
    VMULPS(YMM00, YMM05, YMM05);
    VSUBPS(YMM05, YMM15, YMM15);
  }

  VSTORPD(YMM09, pot[0]);
  VSTORPD(YMM10, xacc[0]);
  VSTORPD(YMM11, yacc[0]);
  VSTORPD(YMM12, zacc[0]);
  VSTORPD(YMM13, xjrk[0]);
  VSTORPD(YMM14, yjrk[0]);
  VSTORPD(YMM15, zjrk[0]);
  accjerk[0].xacc = - 0.125 * (xacc[0] + xacc[2]);
  accjerk[0].yacc = - 0.125 * (yacc[0] + yacc[2]);
  accjerk[0].zacc = - 0.125 * (zacc[0] + zacc[2]);
  accjerk[0].pot  =   0.5   * (pot[0]  + pot[2]);
  accjerk[0].xjrk = - 0.125 * (xjrk[0] + xjrk[1]
                             + xjrk[2] + xjrk[3]);
  accjerk[0].yjrk = - 0.125 * (yjrk[0] + yjrk[1]
                             + yjrk[2] + yjrk[3]);
  accjerk[0].zjrk = - 0.125 * (zjrk[0] + zjrk[1]
                             + zjrk[2] + zjrk[3]);
  accjerk[1].xacc = - 0.125 * (xacc[1] + xacc[3]);
  accjerk[1].yacc = - 0.125 * (yacc[1] + yacc[3]);
  accjerk[1].zacc = - 0.125 * (zacc[1] + zacc[3]);
  accjerk[1].pot  =   0.5   * (pot[1]  + pot[3]);
  accjerk[1].xjrk = - 0.125 * (xjrk[4] + xjrk[5]
                             + xjrk[6] + xjrk[7]);
  accjerk[1].yjrk = - 0.125 * (yjrk[4] + yjrk[5]
                             + yjrk[6] + yjrk[7]);
  accjerk[1].zjrk = - 0.125 * (zjrk[4] + zjrk[5]
                             + zjrk[6] + zjrk[7]);
}
\end{lstlisting}

\section{Accuracy}
\label{sec:accuracy}

In this section, we present the accuracy of our implementation in
terms of errors in accelerations, jerks, and potentials of
individual particles for a given snapshot as well as errors in the
total energy in time integration.

We measure the errors of accelerations, jerks, and potentials of
individual particles in the Plummer's model with $N=1$K, 4K, and 16K
($1{\rm K} = 1024$). Our implementation adopts the standard $N$-body
units, $G=M=r_{\rm v}=1$, where $G$ is the gravitational constant, $M$
and $r_{\rm v}$ are the mass and virial radius of the system,
respectively. The softening length is set to $4/N$. The relative
errors of accelerations, jerks, and gravitational potentials are
defined as $|\bm{a}-\bm{a}_{\rm DP}|/|\bm{a}|$, $|\bm{j}-\bm{j}_{\rm
  DP}|/|\bm{j}|$ and $|\phi-\phi_{\rm DP}|/|\phi|$, respectively,
where the reference values, $\bm{a}_{\rm DP}$, $\bm{j}_{\rm DP}$ and
$\phi_{\rm DP}$ are obtained by computing fully in DP for all
operations, such as computation of distances, square roots, divisions,
etc.

\begin{figure*}
  \begin{center}
    \includegraphics[scale=1.5]{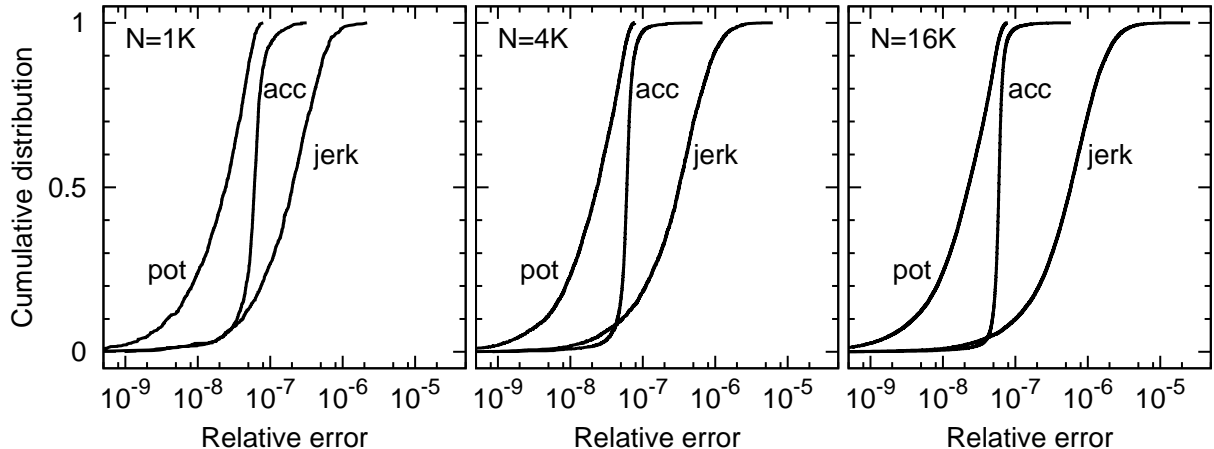}
  \end{center}
  \caption{Cumulative distribution of the relative errors of
    accelerations ("acc"), jerks ("jerk"), and gravitational
    potentials ("pot") calculated in our implementation. The relative
    errors are calculated for the Plummer's model with $N=1$K (left),
    4K (middle), and 16K (right). \label{fig:err_snp}}
\end{figure*}

Figure \ref{fig:err_snp} shows the cumulative distribution of errors
of accelerations, jerks, and gravitational potentials for individual
particles in the Plummer's model with $N=1$K, 4K, and 16K. Most
accelerations and potentials have relative errors around $10^{-8}$.
This is quite natural because, in our implementation, accelerations
and potentials for individual pairs of $i$- and $j$-particles are
calculated in SP. Accuracies of jerks are lower than those of
accelerations and potentials because relative velocity vectors between
$i$- and $j$-particles are calculated in SPs, and their accumulations
are also operated in SP.

\begin{figure}
  \begin{center}
    \includegraphics[scale=1.5]{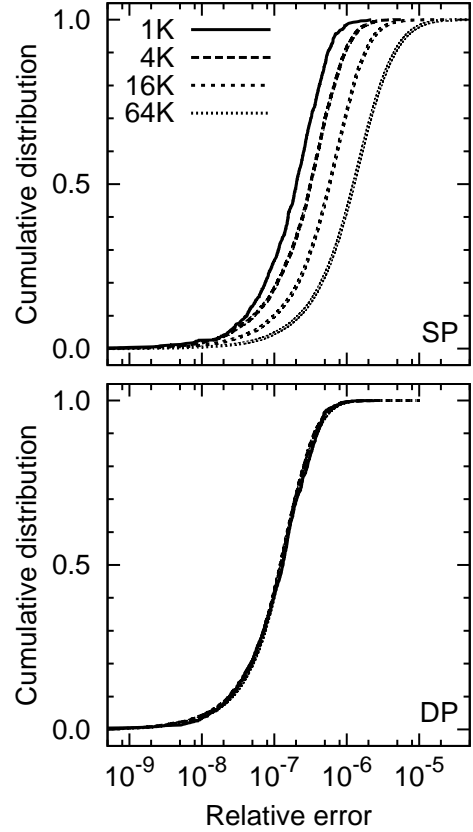}
  \end{center}
  \caption{Cumulative distribution of relative errors of jerks
    accumulated in SP (upper panel) and in DP (lower panel) for
    $N=1$K, 4K, 16K, and 64K.}  \label{fig:err_snp_djerk}
\end{figure}

Note that the relative errors of jerks increase with the number of
particles, which can be clearly seen in the upper panel of figure
\ref{fig:err_snp_djerk}.  This is caused by the errors in accumulating
jerks of individual pairs of $i$- and $j$- particles in SP.  If we
accumulate jerks in DP in the same manner as accelerations and
potentials, the distributions of errors become almost independent of
the number of particle as can be seen in the lower panel of figure
\ref{fig:err_snp_djerk}. We speculate that the origin of the errors
are round-off errors in accumulating jerks in SP. In order to address
the effect of errors in jerks, we perform the two runs with jerks
accumulated in SP, and in DP with various number of particles up to
$N=256$K.
We do not find any significant differences in the energy conservation
throughout the duration of our calculations ($t=0.125$ $N$-body
units). We thus employ the accumulations of jerks in SP, since the
accumulation of jerks in DP degrades the total performance by $20$ --
$25$ \%. However, the accumulations of jerks in SP could lower the
overall accuracy when $N$ is large (say $N > 10^6$).

\begin{figure}
 \begin{center}
  \includegraphics[scale=1.5]{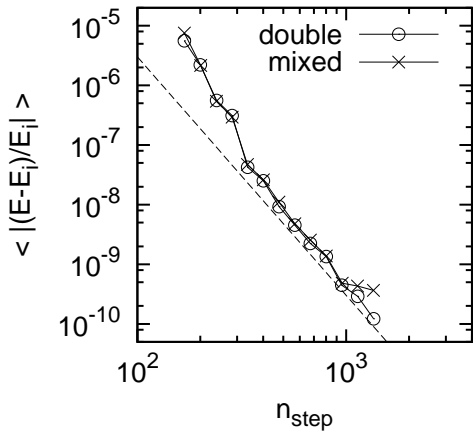}
 \end{center}
 \caption{Averaged relative error in the total energy conservation
   arising from the time integration for $1/8$ crossing time ($0.375$
   $N$-body time units) as a function of the number of steps averaged
   over all particles for $1$ crossing time ($2\sqrt{2}$ $N$-body time
   units), $n_{\rm step}$, in Plummer's model with $N=1K$. The force
   calculations are operated in mixed precision (mixed) and DP
   (double). The dashed line indicates the scaling of $n_{\rm
     step}^{-4}$.}
 \label{fig:err_itg}
\end{figure}

Figure \ref{fig:err_itg} shows relative error in the total energy
conservation averaged over the period of 1/8 crossing time as a
function of the number of timesteps. The number of timesteps, $n_{\rm
  step}$, are averaged over all particles during 1 crossing time,
which corresponds to $2\sqrt{2}$ in the $N$-body unit.  In order to
obtain the errors originated only by the force calculation in the
mixed precisions, we calculate potential energies in DP. The errors in
the total energy conservation decrease with the fourth power of
$n_{\rm step}$, since we adopt the fourth-order Hermite scheme for the
orbit integration of particles. Because of the mixed precision for the
force calculation, the accuracy of the energy conservation stops
improving at the relative errors in the total energy of $10^{-9}$ even
if the number of timesteps increases. On the other hand, the accuracy
of the energy conservation in DP continues to improve for larger
$n_{\rm step}$.

\section{Performance}
\label{sec:performance}

In this section, we present the performance of our implementation of
the collisional $N$-body simulation using the AVX instructions. For
the measurement of the performance, we use Intel Core i7-2600
processors with 8MB cache memory and a frequency of 3.40 GHz, which
contain four processor cores. We disable Intel Turbo Boost Technology
(TB), unless otherwise stated. A compiler which we adopt is
\verb|GCC 4.4.5|. We use compilation options \verb|-O3|,
\verb|-ffast-math|, and \verb|-funroll-loops|. Our code is
parallelized with the Message Passing Interface (MPI) in the same
manner as the NINJA scheme described in NMA06, and we measure the
performance with various numbers of MPI processes on a single
processors.  The code units and softening length are the same as those
in the previous section. We adopt the Plummer's model for the initial
conditions, and evolve the system from time $t=0$ to time $t=1$ in the
$N$-body units.

\begin{figure}
  \begin{center}
    \includegraphics[scale=1.5]{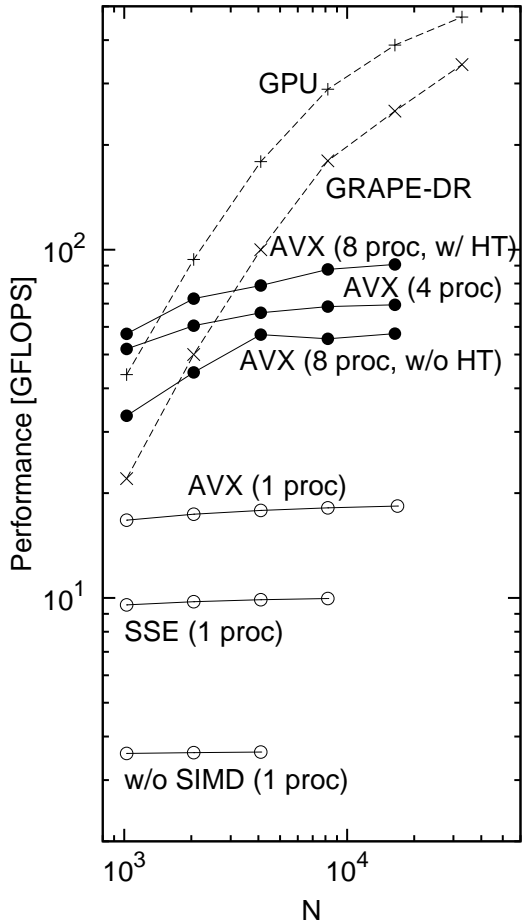}
  \end{center}
  \caption{Performance of our implementation on 1 CPU with $N=1$K, 2K,
    4K, 8K, and 16K. Open and filled circles indicate the results on
    one and four processor cores, respectively. The results with 1, 4
    and 8 MPI processes ("proc") are shown. In the case of 8 MPI
    processes, Hyper-Threading Technology is enabled ("w/ HT"), and
    disabled ("w/o HT"). The results of the implementation only with
    the SSE instructions and with no SIMD instructions are also
    shown. The performances of GRAPE-DR and GPU (GTX 470) are plotted
    in dashed lines for comparison. In order to obtain the performance
    of the GPU, Yebisu $N$-body code \citep{Nitadori09,Spurzem10} is
    used. We use Nitadori's code for the calculation of "w/o
    SIMD". His source code for the force calculation is described in
    List 3 of NMH06.}
  \label{fig:pfm_1cpu}
\end{figure}

Figure~\ref{fig:pfm_1cpu} shows the performance of our implementation
with $N=1$K, 2K, 4K, 8K, and 16K, in which the performance is
calculated by estimating the total calculation cost of an
acceleration, jerk and gravitational potential for a pair of $i$- and
$j$-particles to be $60$ floating-point operations. Although
\cite{Makino01} estimated it to be $57$ floating-point operations, we
adopt the estimate by NMH06 so that one can directly compare the
performance of our implementation with that in NMH06. To evaluate the
performance gain of our implementation with the AVX instructions over
that implemented in C-language without any explicit use of the SSE and
AVX instructions, we measure the performance of the C-language code
shown in the List 3 of NMH06 for $N=1$K, 2K, and 4K on the same
processor. In addition to that, to compare the performance of the AVX
instructions and the SSE instructions, we also develop the
implementation only with the SSE instructions, and measure its
performance for $N=1$K, 2K, 4K and 8K.

The performance of our implementation with the AVX instructions is
$\sim 20$~GFLOPS and higher than that implemented with only the SSE
instructions ($\sim 10$~GFLOPS) by about a factor of two as expected,
and 5 times higher than that implemented without any SIMD instructions
($\sim 4$~GFLOPS). One can see that the performance with 4 MPI
processes on four processor cores reaches $\sim 50$~GFLOPS for $N=1$K
and $\sim 70$~GFLOPS for $N=16$K, which are $2.5$ and $3.5$ times
higher than those with 1 MPI process on one processor core,
respectively. When the Hyper-Threading Technology (HT) is enabled, the
performance with 8 MPI processes on four processor cores reaches
$90$~GFLOPS for $N=16$K, which is higher than that with 4 MPI
processes on four processor cores ($70$~GFLOPS). Note that the
performance with 8 MPI processes on four processor cores ($50$~GFLOPS
for $N=16$K) becomes lower than that with 4 MPI processes on four
processor cores when HT is disabled.

\begin{table}
  \begin{center}
    \caption{Comparison between performances (GFLOPS) with and without
      Intel Turbo Boost Technology (TB) in the cases of 1 MPI process
      on one processor and 8 MPI processes on four processors with the
      Hyper-Threading Technology (HT).}
    \begin{tabular}{|c|c|c|c|c|c|c|c|c|c|c|}
      \hline
      & \multicolumn{5}{|c|}{AVX (1 proc)} & \multicolumn{5}{|c|}{AVX (8 proc, HT)} \\
      \hline
      $N$   & $1$K   & $2$K   & $4$K   & $8$K   & $16$K  & $1$K   & $2$K   & $4$K   & $8$K   & $16$K \\
      \hline
      w/o TB & $16.7$ & $17.4$ & $17.8$ & $18.2$ & $18.4$ & $57.3$ & $72.4$ & $78.9$ & $87.8$ & $90.7$ \\
      \hline
      w/ TB  & $18.7$ & $19.1$ & $19.9$ & $20.2$ & $20.5$ & $58.9$ & $74.5$ & $85.9$ & $91.5$ & $90.9$ \\
      \hline
      Gain  & $1.12$ & $1.10$ & $1.12$ & $1.11$ & $1.11$ & $1.03$ & $1.03$ & $1.09$ & $1.04$ & $1.002$ \\
      \hline
    \end{tabular}
    \label{tab:turbo}
  \end{center}
\end{table}

Table \ref{tab:turbo} shows a comparison between the performance when
TB is enabled and disabled. TB enables CPU frequency to be increased
from $3.4$~GHz to $3.8$~GHz (one active core) and to $3.5$~GHz (four
active cores); thus TB contributes to performance gains of $10$ \%
(one active core) and $3$ \% (four active cores). As expected, one can
see that the gains are, respectively, about $10$ \% and $3$ \% in the
cases of 1 MPI processes on one processor core and 8 MPI processes on
four processor cores.

Note that the performance of our implementation is almost independent
of the number of particles. This feature is advantageous over the
dependences on the number of particles by external hardware
accelerators such as GPUs and GRAPEs. The performances obtained by a
GRAPE-DR board with 4 GRAPE-DR chips ($400$ MHz) mounted and by a GTX
470 GPU with Yebisu code \citep{Nitadori09,Spurzem10} are presented in
figure~\ref{fig:pfm_1cpu} in dashed lines, which strongly depends on
the number of particles and relatively lower for the smaller numbers
of particles.
This is due to the overhead in transferring the data between extended
hardware and host computers. In our implementation with the AVX
instructions, we do not have such an overhead arising from the
transfer of particle data, and thus the performance is almost
independent of the number of particles. One can see that the
performance of our implementation with a single CPU is slightly lower
than that of the GRAPE-DR and GPU systems for $N\lesssim 10^4$ (by
half or one third), and even higher for $N\lesssim 1\times 10^3$. This
feature is quite desirable for simulations on massively parallel
computers, because the performance independent of the number of
particles allows us to achieve fairly good ``strong
scaling''. Furthermore, this feature is also advantageous under wide
dynamic range, such as the case of the presence of binaries in star
clusters.

\section{Summary and discussion}
\label{sec:summary}

We have developed an $N$-body code for a collisional system with the
new SIMD instruction set available on the Sandy Bridge architecture
released by Intel Corporation, the AVX instruction set. The
performance is $\sim 20$ GFLOPS with one processor core, which is two
times higher than that with the SSE instructions as expected, and five
times higher than that in C-language without any explicit use of SIMD
instructions. The performance on a single CPU with four processor
cores reaches $90$~GFLOPS.

Here, let us estimate the performance of our code parallelized with
the NINJA scheme on massively parallel systems. As a reference, NMA06
achieved about $4$ GFLOPS with one processor core in a dual-core
Opteron processors, and $2.03$ tera FLOPS (TFLOPS) with $400$
dual-core Opteron processors for an $N=64$K system using the NINJA
scheme. Since we have achieved $20$ GFLOPS with one core on a Intel
Core i7-2600 processor, the performance of 10~TFLOPS can be achieved
with a massively parallel system with the same number of processor
cores of Core-i7 2600 processors.

There are two advantages of our approach to accelerate $N$-body
simulations using SIMD instructions implemented on CPUs over the use
of external hardwares such as GPUs and GRAPE families. One is the
portability of our code. We can carry out $N$-body simulations with
good performance on any computer systems without GPUs and GRAPE
boards. Even on the systems with instruction set architectures other
than the x86 architecture, the fact that SIMD instruction sets similar
to the SSE and AVX instruction set are available (e.g. Vector
Multimedia Extension on Power series, HPC-ACE on SPARC64 VIIIfx, etc.)
suggests that our approach to use SIMD instructions is successful on
any architecture. The second advantage is the fact that the
performance of our implementation is almost independent of the number
of particles because there is no overhead associate with the
communication between a CPU and an external hardware, such as a GRAPE
and a GPU. As a result, the performance of our implementation is
better than that aided by external hardwares. In other words, our code
have very good strong scaling on massively parallel systems.

Finally, let us mention the future improvement of our implementation.
The performance of our $N$-body code will be higher when Fused
Multiply-Add (FMA) instructions are introduced. Using the FMA
instructions which compute the product of two numbers and add the
result to another number in one step, we will be able to improve the
performance and the accuracy of the code, especially accumulation of
jerks in the force loop. The FMA instructions will be available in the
next-generation CPU named "Bulldozer" by AMD Corporation, and
"Haswell" by Intel Corporation, which are scheduled to be released in
mid-2011, and in 2013, respectively. Our code is publicly available at
http://code.google.com/p/phantom-grape/.

\section*{Acknowledgments}

We thank Junichiro Makino for providing the performance data of
GRAPE-DR. Numerical simulations have been performed with computational
facilities at the Center for Computational Sciences in the University
of Tsukuba. This work was supported by Scientific Research for
Challenging Exploratory Research (21654026), Grant-in-Aid for Young
Scientists (start-up: 21840015), the FIRST project based on the
Grants-in-Aid for Specially Promoted Research by MEXT (16002003), and
Grant-in-Aid for Scientific Research (S) by JSPS (20224002).

\end{document}